\documentclass[final, 3p, times, twocolumn]{elsarticle}

\usepackage{amssymb, amsmath, amsthm, bm}
\usepackage{geometry}
\usepackage{algorithm}
\usepackage{algpseudocode}
\usepackage{empheq}
\usepackage{graphicx}
\graphicspath{{figures-pdf//}}
\usepackage[normalem]{ulem}
\usepackage{url}
\usepackage{cases}
\usepackage{subeqnarray}
\usepackage{bm}
\usepackage{graphicx, color, overpic}
\usepackage{float}
\usepackage{booktabs}
\usepackage{diagbox}
\usepackage{multirow}
\newlength\figsep
\usepackage{enumitem}
\newlength\FourImW
\newlength\BigOneImW
\newlength\twofigwidth
\newlength\ThreeImW
\newlength\SixImW
\newlength\sfigwidth
\newcommand{\Dec}{\operatorname{Dec}}
\newcommand{\Spl}{\operatorname{Spl}}

\journal{IEEE Multimedia}
\usepackage{lineno}

\newtheorem{Property}{Property}
\newtheorem{Fact}{Fact}

\usepackage[font=small, skip=0pt]{caption}
\usepackage[bookmarks=false]{hyperref}
\hypersetup{
 linktocpage=true, 
 pdfborderstyle={/S/S/W 1}, 
 hyperindex=true, 
 bookmarks=false, 
 bookmarksopen=false, 
 bookmarksnumbered=false, 
}

\begin{document}

\begin{frontmatter}

\title{Cryptanalyzing an Image Encryption Algorithm Underpinned by 2{D} Lag-Complex {L}ogistic Map}

\author[xtu-cn]{Chengqing Li\corref{corr}}
\ead{DrChengqingLi@gmail.com}

\author[ICIP-cn]{Xianhui Shen}

\author[hnu-cn]{Sheng Liu}

\cortext[corr]{Corresponding author.}

\address[xtu-cn]{School of Computer Science, Xiangtan University, Xiangtan 411105, Hunan, China}

\address[ICIP-cn]{MOE (Ministry of Education) Key Laboratory of Intelligent Computing and Information Processing, Xiangtan University, Xiangtan 411105, China}

\address[hnu-cn]{School of Computer Science and Electronic Engineering, Hunan University, Changsha 410082, Hunan, China}

\begin{abstract}
This paper analyzes security performance of an image encryption algorithm using 2D lag-complex Logistic map (LCLM), which adopts it as a pseudo-random number generator, and uses the sum of all pixel values of the plain-image as its initial value to control the random combination of the basic encryption operations. However, multiple factors make the final 
pseudo-random sequences controlling the encryption process may be the same for different plain-images. Based on this point, we proposed a chosen-plaintext attack by attacking the six encryption steps with a strategy of divide and conquer. Using the pitfalls of 2D-LCLM, the number of required chosen plain-images is further reduced to $5\cdot\log_2(MN)+95$, where $\mathit{MN}$ is the number of pixels of the plain-image.
\end{abstract}
\end{frontmatter}

\section{Introduction}

With the popularization of multimedia capture devices and the mobile Internet, more and more people become enjoying sharing and transmitting photos via a social platform, such as WeChat, Instagram and Facebook.
Meanwhile, due to the openness of the network, multimedia information
itself or the contained privacy may be leaked during the transmission process~\cite{Balancing:NDSS19}. Such underlying threats are transparent for most people. Therefore, ensuring security and privacy of multimedia data, especially that of digital images, with proper balancing point with usability has become needs of everyone living with cyberspace.
Due to the special characteristics of image data, such as the large amount of data itself and strong correlation existing between neighbouring pixels, it is generally impractical to encrypt the protected object with the classic text encryption algorithms by converting the object into one-dimensional text data.
Under this background, a large number of image encryption algorithms were proposed in the past three decades~\cite{ye2016image, zhouli:synch:SC21, wuxj:BCNN:IM22}. Unfortunately, some were found insecure of different extents from the perspective of modern cryptanalysis \cite{Lim:order:IM18, cqli:block:JISA20, chenjx:CPA:TM22, Lingfeng:hiding:TM22, cqli:delay:IEEEM22, cqli:DNA:JVCI22}. With the development of neural network technology, the form of attack and defense applied on image data has been changed much~\cite{wuxj:BCNN:IM22, zhou:CS:TM21, cqli:CS:TMM22}.

As a cryptosystem can be considered as a complex system defined on a digital domain, various nonlinear sciences, e.g. chaos, synchronization, were adopted as alternative methodologies to design new image encryption algorithms \cite{Lim:order:IM18, ping:ps:NC18, Hua2021ND, zhang:spatio:NCA20}.
However, the nice complex dynamics demonstrated in the original infinite-precision domain may be
degenerated to a certain extent and even diminished in a finite-precision arithmetic domain \cite{min:Oscillator:IJBC21,cqli:network:TCASI2019}.
As shown in \cite{cqli:block:JISA20, cqli:DNA:JVCI22}, the randomness of the sequence obtained by iterating a discrete chaotic map on a computer with finite arithmetic precision is much lower than that expected by the designers: the periods of the sequences starting from some initial conditions are even smaller than three (see Fig.~\ref{fig:network}).

In \cite{zhangf:2Dlag:IM21}, 2D lag-complex Logistic map (2D-LCLM) is designed to enhance dynamic complexity of the map and counteract its dynamic degradation in digital domain.
Then, an image encryption algorithm based on the special Logistic map (IEALM) is designed and tested to demonstrate cryptographic application merits of the map. In the algorithm, the pseudo-random number sequences generated by iterating 2D-LCLM are used to obfuscate and diffuse the plain-image via complex cascade of modulo addition, XOR operation, and bit-level permutation.
This paper reevaluated security performance of IEALM and reported the following points: 1)
the correlation mechanism between the encryption process and the plain-image is invalid with a non-negligible probability; 2) some intrinsic properties of 2D-LCLM are disclosed to support
a chosen-plaintext attack; 3) the basic encryption parts can be separately attacked with some chosen plain-images in turn; 4) experimental results and complexity analysis of the attack are provided to demonstrate its effectiveness.

The rest of this paper is organized as follows.
Section~\ref{sec:IEALM} briefly introduces the procedure of IEALM.
Section~\ref{sec:cryptanalysis} presents the insecurity metrics of 2D-LCLM, and
the security vulnerabilities of IEALM, together with some experimental results.
The last section concludes the paper.

\section{Description of IEALM}
\label{sec:IEALM}

The encrypted object of IEALM is an 8-bit RGB image of size $M\times N\times 3$.
The algorithm does not consider the specific storage format of the plain-image, 
and directly splits it into three separate color channels, which are then processed by the same encryption process.
For brevity, we describe the encryption procedure for only one channel.
The plain-image is scanned in the raster order and then denoted by $\bm{I}=\{I(i)\}_{i=0}^{MN-1}$.
Likewise, the corresponding cipher-image is represented as $\bm{I}''=\{I''(i)\}_{i=0}^{MN-1}$.
Then, IEALM can be described as follows.
\begin{itemize}[leftmargin=*]
\item \textit{The secret key}: two control parameters of 2D-LCLM
\begin{equation}\label{eq:LCLM}
\left\{
\begin{aligned}
x(i+1) &=b\cdot x(i)\cdot (1-z(i)), \\
y(i+1) &=b\cdot y(i)\cdot (1-z(i)), \\
z(i+1) &=a\cdot x(i)^2+y(i)^2, 
\end{aligned}
\right.
\end{equation}
and its two distinct initial conditions $K_1=(x(0), y(0), z(0))$ and $K_2=(x'(0), y'(0), z'(0))$, where $a=2$ and $b\in[1.69, 2)$.

\item \textit{Initialization}:

\begin{itemize}[leftmargin=*]
\item \textit{Step 1}:
Calculate two initial conditions through
\begin{equation}
\label{eq:plaintokey}
\left\{
\begin{aligned}
K_1 &=(0.2+X_r/10^9, 0.4+Y_g/10^9, 0.1+Z_b/10^9), \\
K_2 &=(0.3+X_r/10^9, 0.5+Y_g/10^9, 0.2+Z_b/10^9), 
\end{aligned}
\right.
\end{equation}
where $X_r$, $Y_g$, and $Z_b$ are the sum of the pixel values of 
red, green, and blue channels of the plain-image, respectively.

\item \textit{Step 2}: Iterate Eq.~\eqref{eq:LCLM} $2MN+250$ times from the initial condition $K_1$ and discard the first 250 elements to get three chaotic sequences $\bm{X}=\{x(i)\}_{i=0}^{2MN-1}$, $\bm{Y}=\{y(i)\}_{i=0}^{2MN-1}$ and $\bm{Z}=\{z(i)\}_{i=0}^{2MN-1}$.
Then, generate a sequence $\bm{G}=\{g(i)\}_{i=0}^{2MN-1}$, where
\begin{equation*}
g(i)=\frac{x(i)+y(i)+z(i)}3.
\end{equation*}

\item \textit{Step 3}: Generate a $4$-bit integer sequence $\bm{U}=\{U(i)\}_{i=0}^{MN-1}$ and two $8$-bit integer sequences $\bm{V}=\{V(i)\}_{i=0}^{MN-1}$ and $\bm{W}=\{W(i)\}_{i=0}^{MN-1}$ via
\begin{equation*}
\left\{
\begin{aligned}
U(i) &= \left\lfloor |x(i)| \cdot 10^{15} \right\rfloor \bmod 16, \\
V(i) &= \left\lfloor \Dec(y(i)) \cdot 10^3\right\rfloor\bmod 256, \\
W(i) &= \left\lfloor \Dec(z(i)) \cdot 10^3\right\rfloor\bmod 256, 
\end{aligned}
\right.
\end{equation*}
where $\Dec(x)=x\cdot 10^3-\lfloor x\cdot 10^3 \rfloor$.

\item \textit{Step 4}: Generate two permutation vectors $\bm{T}_{1, 0}=\{T_{1, 0}(i)\}_{i=0}^{MN-1}$ and $\bm{T}_{2, 0}=\{ T_{2, 0}(i)\}_{i=0}^{MN-1}$, where $T_{1, 0}(i)$ and $T_{2, 0}(i)$
are the order of $X(i)$ and $X(i+MN)$ in sets $\left\{X(i)\right\}_{i=0}^{MN-1}$ and $\{X(i)\}_{i=MN}^{2MN-1}$,
respectively.
Similarly, obtain six permutation vectors of length $\mathit{MN}$, $\bm{T}_{1, 1}$, $\bm{T}_{2, 1}$, $\bm{T}_{1, 2}$, $\bm{T}_{2, 2}$, $\bm{T}_{1, 3}$ and $\bm{T}_{2, 3}$, from $\bm{Y}$, $\bm{Z}$, and $\bm{G}$, respectively.
Group the above vectors as $\bm{T}_1=\left\{\bm{T}_{1, 0}, \bm{T}_{1, 1}, \bm{T}_{1, 2}, \bm{T}_{1, 3}\right\}$ and $\bm{T}_2=\left\{\bm{T}_{2, 0}, \bm{T}_{2, 1}, \bm{T}_{2, 2}, \bm{T}_{2, 3}\right\}$.

\item \textit{Step 5}: Repeat \textit{Step 2, 3} and \textit{4} using initial condition $K_2$ and produce the counterparts $\bm{X}'$, $\bm{Y}'$, $\bm{Z}'$, $\bm{G}'$, $\bm{U}'$, $\bm{V}'$, $\bm{W}'$, $\bm{T}_3=\left\{\bm{T}_{3, 0}, \bm{T}_{3, 1}, \bm{T}_{3, 2}, \bm{T}_{3, 3}\right\}$ and $\bm{T}_4=\left\{\bm{T}_{4, 0}, \bm{T}_{4, 1}, \bm{T}_{4, 2}, \bm{T}_{4, 3}\right\}$.
\end{itemize}

\item \textit{The encryption procedure}:

\begin{itemize}[leftmargin=*]
\item \textit{Step 1}: Perform modulo addition on $\bm{I}$ to obtain a sequence $\bm{I}^*=\{I^*(i)\}_{i=0}^{MN-1}$ by
\begin{equation}
\label{eq:P*}
I^*(i) = I(i)\boxplus V(i), 
\end{equation}
where $a\boxplus b=(a+b)\bmod 2^{n_0}$ and $n_0$ is the binary length of $a$ and $b$.

\item \textit{Step 2}: Perform XOR operation on $\bm{I}^*$ to obtain sequence $\bm{I}^{**}=\{I^{**}(i)\}_{i=0}^{MN-1}$ by
\begin{equation}
\label{eq:Pj}
I^{**}(i) = W(i)\oplus I^*(i).
\end{equation}

\item \textit{Step 3}: Divide $\bm{I}^{**}$ into two 4-bit sequences
$\bm{L}^{**}=\{L^{**}(i)\}_{i=0}^{MN-1}$, $\bm{H}^{**}=\{H^{**}(i)\}_{i=0}^{MN-1}$ with function $\Spl(\bm{X})$ that returns two sequences $\bm{X}_L=\{X(i)\bmod 16\}_{i=0}^{MN-1}$ and $\bm{X}_H=\{
\lfloor X(i)/16 \rfloor \}_{i=0}^{MN-1}$, where
$\bm{X}=\{X(i)\}_{i=0}^{MN-1}$.

\item \textit{Step 4}: Perform bit-level permutation on $\bm{L}^{**}$ and $\bm{H}^{**}$ to get sequences
$\widetilde{\bm{L}}^{**}=\{\widetilde{L}^{**}(i)\}_{i=0}^{MN-1}$
and
$\widetilde{\bm{H}}^{**}=\{\widetilde{H}^{**}(i)\}_{i=0}^{MN-1}$:
$\widetilde{L}^{**}(i)=\sum_{k=0}^3L^{**}_k(T_{1, k}(i))\cdot2^k$
and $\widetilde{H}^{**}(i)=\sum_{k=0}^3H^{**}_k(T_{2, k}(i))\cdot2^k$.
Then produce sequence $\bm{L}'$ by
\begin{equation}
\label{eq:L'}
L'(i)=U(i)\oplus\widetilde{L}^{**}(i)\oplus\widetilde{H}^{**}(i).
\end{equation}

\item \textit{Step 5}: Perform bit-level permutation on $\bm{L}'$ and $\bm{H}^{**}$ to get sequences
$\widehat{\bm{L}}'=\{\widehat{L}'(i)\}_{i=0}^{MN-1}$
and
$\widehat{\bm{H}}^{**}=\{\widehat{H}^{**}(i)\}_{i=0}^{MN-1}$:
$\widehat{L}'(i)=\sum_{k=0}^3L'_k(T_{3, k}(i))\cdot2^k$
and $\widehat{H}^{**}(i)=\sum_{k=0}^3H^{**}_k(T_{4, k}(i))\cdot2^k$,
where $L'(i)=\sum_{k=0}^3L'_{k}(i)\cdot2^k$.
Then produce sequence $\bm{H}'=\{H'(i)\}_{i=0}^{MN-1}$ via
\begin{equation}
\label{eq:H'}
H'(i)=U'(i)\oplus\widehat{L}'(i)\oplus\widehat{H}^{**}(i).
\end{equation}

\item \textit{Step 6}: Combine two 4-bit intermediate sequences $\bm{L}'$ and $\bm{H}'$ into one 8-bit sequence
$\bm{I}'=\{I'(i)\}_{i=0}^{MN-1}=\{L'(i)+H'(i)\cdot16\}_{i=0}^{MN-1}$.
Then perform further confusion on $\bm{I}'$ to obtain cipher-image $\bm{I}''$, where
\begin{equation}
\label{eq:C}
I''(i)=W'(i)\oplus(I'(i)\boxplus V'(i)).
\end{equation}
\end{itemize}

\end{itemize}

\section{Cryptanalysis of IEALM}
\label{sec:cryptanalysis}

The designers of IEALM emphasized that it can ``well resist the chosen-plaintext attack and other classical attacks. Each image corresponds to a different secret key, so it has high security and can resist plaintext attacks"~\cite{zhangf:2Dlag:IM21}.
In this section, we dispute such claim based on the intrinsic properties of 2D-LCLM and IEALM.

\subsection{The properties of 2D-LCLM}

IEALM uses 2D-LCLM as a pseudo-random number generator to control the encryption process. According to Property~\ref{pro:xycontact} and \ref{pro:rank}, one can see that four equations on the variables of IEALM, $\bm{T}_{1, 0}=\bm{T}_{1, 1}$, $\bm{T}_{2, 0}=\bm{T}_{2, 1}$, $\bm{T}_{3, 0}=\bm{T}_{3, 1}$ and $\bm{T}_{4, 0}=\bm{T}_{4, 1}$ always hold, which make the algorithm is more vulnerable to chosen-plaintext attack.

\begin{Property}
\label{pro:xycontact}
The coordinates of 2D-LCLM, $x(i)$, $y(i)$, satisfy $x(i)/y(i)=x(j)/y(j)$ for any $i, j \in \{0, 1, \cdots, MN-1\}$.
\end{Property}
\begin{proof}
If $i=j$, it is obvious that $x(i)/y(i)=x(j)/y(j)$.
Assuming $i>j$, one has
\begin{align}
\label{eq:xiexp}
 x(i) &= bx(i-1) (1-z(i)) \notag \\
 &= b^{i-j-1}x(j) (1-z(i))\cdots (1-z(j+1)) \notag \\
 &= b^{i-j-1}x(j)\prod_{t=j+1}^{i} (1-z(t)).
\end{align}
Similarly, one can get
 \[
 y(i)=b^{i-j-1}y(j)\prod_{t=j+1}^{i} (1-z(t)).
\]
Dividing the two sides of Eq.~\eqref{eq:xiexp} with that of the above equation, one can get $x(i)/y(i)=x(j)/y(j)$.
\end{proof}

\begin{Property}
\label{pro:rank}
If the coordinates of $\bm{X}=\{x(i)\}_{i=0}^N$ and $\bm{Y}=\{y(i)\}_{i=0}^N$ satisfy $x(i)/y(i)=x(j)/y(j)$, where $i, j\in\{0, 1, \cdots, N\}$. Then the ranking order of $x(i)$ in $\bm{X}$ is the same as that of $y(i)$ in $\bm{Y}$.
\end{Property}
\begin{proof}
According to property~\ref{pro:xycontact}, it can be known that
\begin{equation*}
y(j)=A(j)y(i), 
\end{equation*}
where $A(j)=x(j)/x(i)$ and $j=0, 1, \cdots, N$.
Let $\bm{A}=\{A(j)\}_{j=0}^N$, and then according to the known condition, one can know there are $k-1$ elements in $\bm{A}$ that are less than $1$. Then one can deduce that there are $k-1$ elements in $\bm{Y}$ that are smaller than $y(i)$.
Namely, $y(i)$ is ranked $k$ in $\bm{Y}$.
\end{proof}

\begin{figure*}[!htb]
\centering
	\begin{minipage}[t]{\ThreeImW}
	\centering
	\includegraphics[width=\ThreeImW]{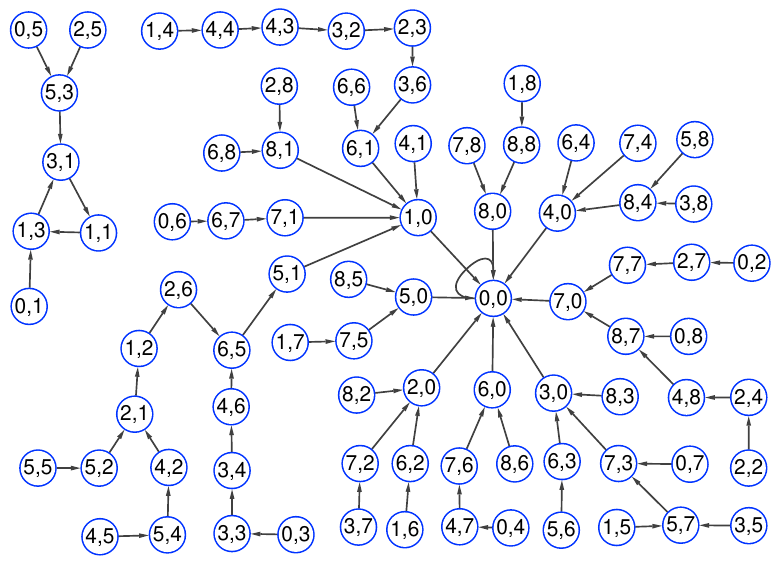}
	a)
	\end{minipage}
	\begin{minipage}[t]{\ThreeImW}
	\centering
	\includegraphics[width=\ThreeImW]{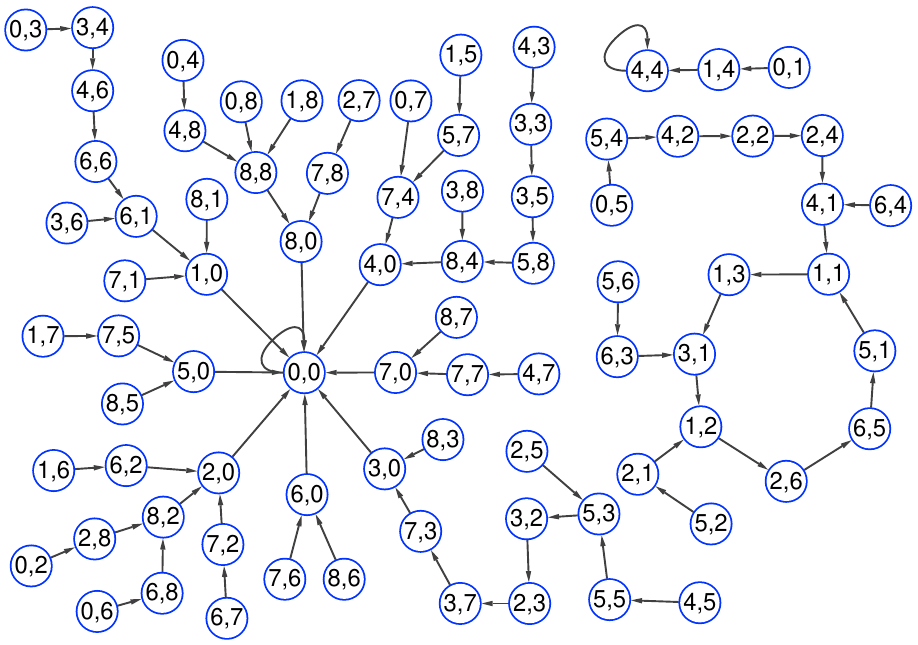}
	b)
	\end{minipage}
	\begin{minipage}[t]{\ThreeImW}
	\centering
	\includegraphics[width=\ThreeImW]{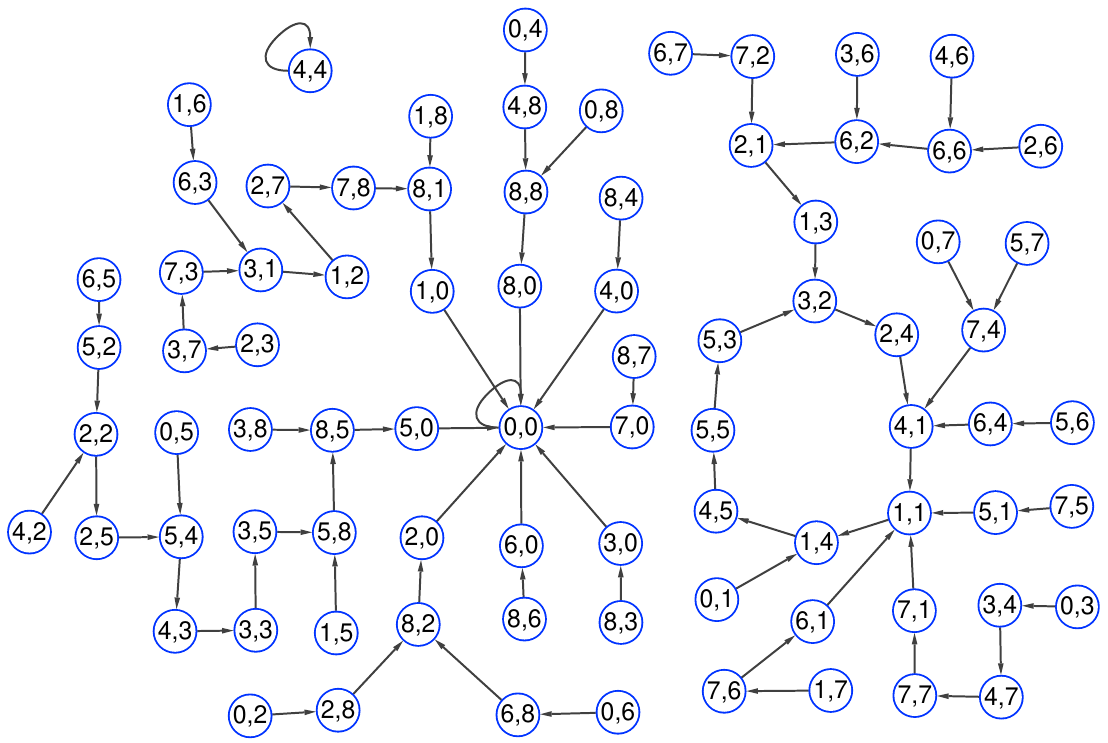}
	c)
	\end{minipage}
\caption{The functional graph of the third equation in Eq.~\eqref{eq:newLCLM} with $b=511/2^8$ under 3-bit fixed-point precision and different quantization strategies: a) floor; b) round; c) ceil, where symbol $(i, j)$ in each node denotes coordinate $(i/2^3, j/2^3)$.}
\label{fig:network}
\end{figure*}

\begin{Property}
\label{col:newmap}
2D-LCLM can be represented as
\begin{equation}
\label{eq:newLCLM}
\left\{
\begin{aligned}
 x(i+1) &= bx(i)(1-(a+(x(0)/y(0))^2)x(i-1)^2), \\
 y(i+1) &= by(i)(1-(a(y(0)/x(0))^2+1)y(i-1)^2), \\
 z(i+1) &= b^2z(i)(1-z(i-1))^2, 
\end{aligned}
\right.
\end{equation}
where $i\ge 1$, $(x(0), y(0), z(0))$ is the initial condition, $x(1)=bx(0)(1-z(0))$, $y(1)=by(0)(1-z(0))$, and $z(1)=ax(0)^2+y(0)^2$.
\end{Property}
\begin{proof}
According to Eq.~\eqref{eq:LCLM} and Property~\ref{pro:xycontact}, when $i\geq 1$, one has
\begin{align*}
 x(i+1) &=bx(i)(1-ax(i-1)^2-y(i-1)^2) \\
   &=bx(i)(1-(a+(y(0)/x(0))^2)x(i-1)^2).
\end{align*}
Likewise, one can get
\begin{align*}
 y(i+1)=by(i)(1-(a(x(0)/y(0))^2+1)y(i-1)^2).
\end{align*}
According to Eq.~\eqref{eq:LCLM}, when $i\geq 1$, one can obtain
\begin{align*}
 z(i+1) &= ax(i)^2+y(i)^2 \\
   &= b^2 (ax(i-1)^2+y(i-1)^2) (1-z(i-1))^2 \\
   &= b^2z(i) (1-z(i-1))^2.
\end{align*}
\end{proof}

The designers of IEALM stated ``2D-LCLM is 3D in a real field" in \cite{zhangf:2Dlag:IM21}, saying that it can show better properties than the other 2D chaotic maps.
Unfortunately, Property~\ref{col:newmap} negates the statement, as it shows that three maps of 2D-LCLM are independent, and each independent map can not traverse the whole 3D domain.

To disclose the real structure of 2D-LCLM in a computer, we performed comprehensive tests on its functional graphs as \cite{cqli:delay:IEEEM22, cqli:DNA:JVCI22, cqli:network:TCASI2019}.
Three typical examples are depicted in Fig.~\ref{fig:network}.
Any orbit definitely enter a cycle after a transient process. Note that
this rule always keep no matter how large the implementation precision is. As shown in \cite{cqli:delay:IEEEM22, cqli:network:TCASI2019}, some cycles of short period (even self-loop) always exist no matter which enhancement method is adopted, e.g.
increasing the arithmetic precision, perturbing states, perturbing the control parameters, 
switching among multiple chaotic maps and cascading more than one chaotic maps.
If the initial state is located in a small-scale connected component or a cycle of short period in the functional graph of the used chaotic map, there are even no enough available states to be discarded.
So, an adaptive threshold should be set to avoid this problem. But, it would cost additional computation.

\subsection{Key distribution}

To thwart known/chosen-plaintext attack, IEALM uses the information of plain-image to generate different secret keys when encrypting different plain-images, but the uniformity of key distribution is ignored.
For an encryption algorithm, it is secure if the keys obey a uniform distribution, because all keys have the same probability of being selected.
A classical example of insecurity is substitution cipher (e.g. Caesar cipher), which uses a codebook to directly replace one letter with another.
Since the frequency of letters does not obey a uniform distribution, the algorithm cannot resist statistical attack.

\begin{figure}[!htb]
\centering
\includegraphics[width=\BigOneImW]{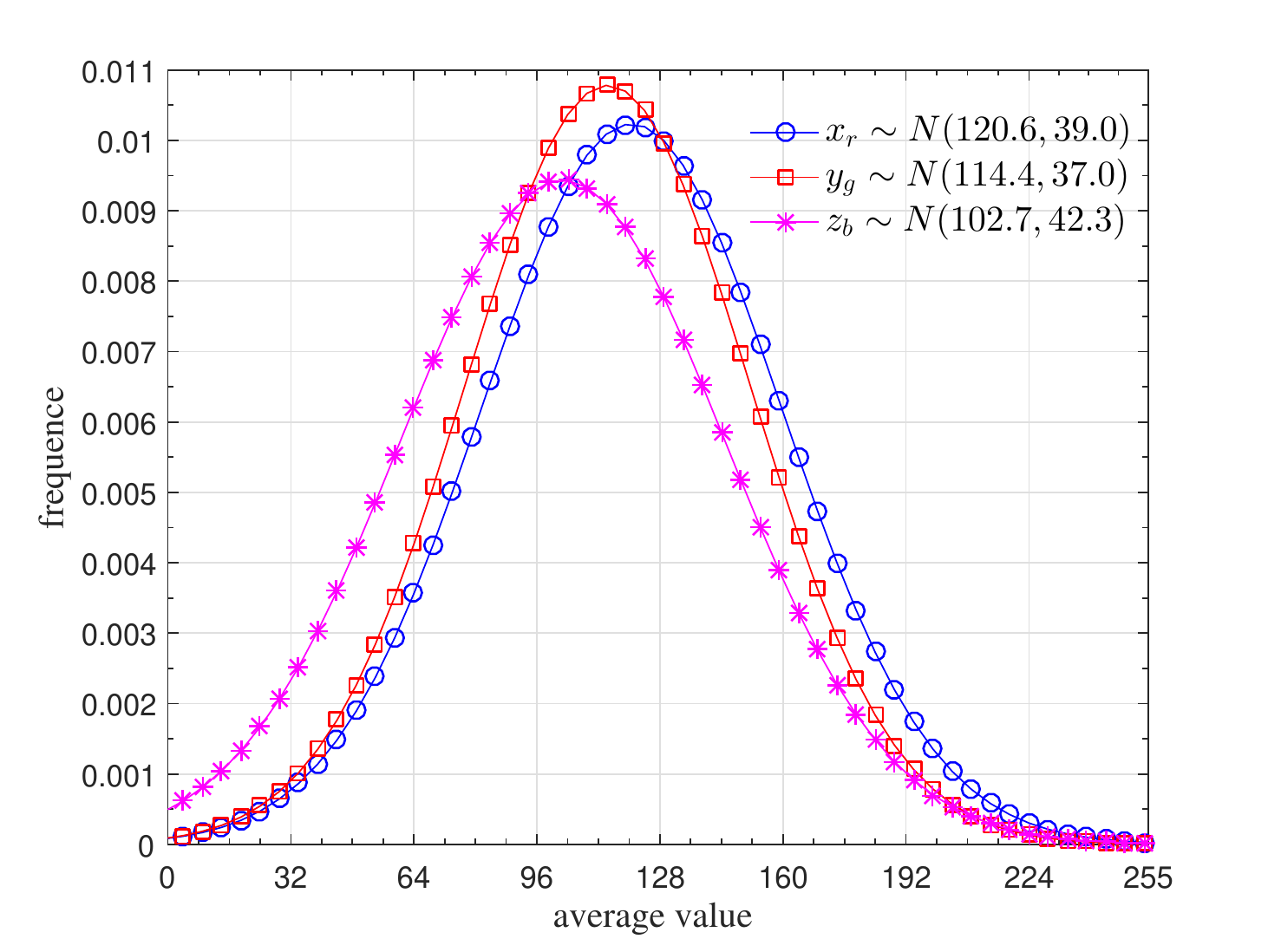}
\caption{The distribution of the average pixel value of images of Mini-ImageNet, where $x_r$, $y_g$, and $z_b$ are the average pixel values of red, green, and blue channels of images, respectively.}
\label{fig:Distribution}
\end{figure}

In IEALM, the sum of pixel value of the plain-image is used to generate the initial value of 2D-LCLM.
Since a natural image has its practical visual meaning, the sum is not random and does not obey a uniform distribution.
To show such property of natural images, we tested the distribution of the average pixel value of 60,000 images of Mini-ImageNet \cite{Vinyals:NIPS2016} and found that the value follows a normal distribution, as shown in Fig.~\ref{fig:Distribution}.

According to the above experimental conclusion, if we arbitrarily select a $256\times 256$ image from Mini-ImageNet, in theory, a total of $4.7\cdot 10^{21}$ possible keys can be generated.
However, since the mean pixel values are normally distributed, the probabilities of these keys being selected are not equal.
According to the property of the normal distribution, one can get $\mathrm{Prob}(81.641<x_r<159.609)=0.6827$, $\mathrm{Prob}(77.388<y_g<151.382)=0.6827$ and $\mathrm{Prob}(60.422<z_b<144.984)=0.6827$.
As for an attacker, only by enumerating this part of the secret key through a brute force attack, the success probability for attacking the algorithm is at least 30\%.
However, this part accounts for only 2.91\% of the total key space.

\subsection{Chosen-plaintext Attack}

In \cite{zhangf:2Dlag:IM21}, the designers adopted the sum of pixel values of each plain-image as the secret key, trying to achieve the function of ``one-time passwords". But it is also possible to generate the same key for different plain-images. Firstly, the same key must be generated for two identical plain-images. Secondly, for different plain-images it may also generate the same key, such as different images with the same sum of pixel values. In addition, due to the limited arithmetic precision of computer, even different plain-images owning
different average pixel values may generate the same secret key through the calculation of Eq.~\eqref{eq:plaintokey}.

As for plain-image $\bm{I}_0$ and its corresponding cipher-image $\bm{I}_0''$, one can try to decrypt another cipher-image $\bm{I}_1''$ if the following two conditions hold at the same time: 1) $\bm{I}_1''$ is encrypted with the same secret key as $\bm{I}_0''$; 2) the corresponding plain-image of $\bm{I}_1''$ has the same result calculated with Eq.~\eqref{eq:plaintokey} as $\bm{I}_0$, namely the final sequences controlling the encryption process are the same.

\begin{itemize}[leftmargin=*]
\item Simplification of IEALM

For simplicity, let $(\bm{W}_L, \bm{W}_H)=\Spl(\bm{W})$, $(\bm{W}_L', \bm{W}_H')=\Spl(\bm{W}')$, $(\bm{V}_L, \bm{V}_H)=\Spl(\bm{V})$, $(\bm{V}_L', \bm{V}_H')=\Spl(\bm{V}')$, and $(\bm{L}^*$, $\bm{H}^*)=\Spl(\bm{I}^*)$.
From Eq.~\eqref{eq:L'}, as for the four lower bit-planes, one has
\begin{align}
\label{eq:newL'}
L'(i)
&=U(i)\oplus\widetilde{W}_L(i)\oplus\widetilde{W}_H(i)\oplus\widetilde{L}^*(i)\oplus\widetilde{H}^*(i) \notag \\
&=\beta(i)\oplus L^{\star}(i), 
\end{align}
where 
\begin{equation*}
L^{\star}(i)=\widetilde{L}^*(i)\oplus\widetilde{H}^*(i),
\end{equation*}
$\beta(i)=U(i)\oplus\widetilde{W}_L(i)\oplus\widetilde{W}_H(i)$,
$\widetilde{W}_L(i)=\sum_{k=0}^3 W_{L, k}(T_{1, k}(i))\cdot2^k$, 
$\widetilde{W}_H(i)=\sum_{k=0}^3 W_{H, k}(T_{2, k}(i))\cdot2^k$, 
$\widetilde{L}^*(i)=\sum_{k=0}^3 L_k^*(T_{1, k}(i))\cdot2^k$, 
and $\widetilde{H}^*(i)=\sum_{k=0}^3 H_k^*(T_{2, k}(i))\cdot2^k$.
According to Eq.~\eqref{eq:H'}, for the four higher bit-planes, one has
\begin{align}
\label{eq:newH'}
H'(i)
& = U'(i)\oplus \widehat{L}'(i)\oplus \widehat{W}_H(i)\oplus \widehat{H}^*(i) \notag\\
& = U'(i)\oplus \widehat{\beta}(i) \oplus \widehat{L}^{\star}(i)\oplus \widehat{W}_H(i)\oplus \widehat{H}^*(i) \notag\\
& = \beta'(i)\oplus H^{\star}(i), 
\end{align}
where
\begin{equation*}
H^{\star}(i)=\widehat{L}^{\star}(i)\oplus \widehat{H}^*(i),
\end{equation*}
$\beta'(i)=U'(i)\oplus \widehat{\beta}(i)\oplus \widehat{W}_H(i)$,
$\widehat{\beta}(i)=\sum_{k=0}^3\beta_k(T_{3, k}(i))\cdot2^k$, 
$\widehat{W}_H(i)=\sum_{k=0}^3W_{H, k}(T_{4, k}(i))\cdot2^k$, 
$\widehat{L}^{\star}(i)=\sum_{k=0}^3L^{\star}(T_{3, k}(i))\cdot2^k$, 
and $\widehat{H}^*(i)=\sum_{k=0}^3H_k^*(T_{4, k}(i))\cdot2^k$.

\begin{Fact}
\label{fact:modxorhighbit}
$\forall\ a, b\in \{0, 1, \cdots, 2^{n_0}-1\}$, $(a\oplus 2^{n_0-1})\boxplus b=(a\boxplus
b)\oplus 2^{n_0-1}=a\oplus (b\boxplus 2^{n_0-1})$.
\end{Fact}

According to Eq.~\eqref{eq:P*}, Eq.~\eqref{eq:Pj} and Fact~\ref{fact:modxorhighbit}, one can get
\begin{align*}
I^{**}(i) &= W(i)\oplus(I(i)\boxplus V(i)) \\
   &= (W(i) \oplus 128)\oplus(I(i)\boxplus (V(i) \oplus 128)), 
\end{align*}
where $i=0, 1, \cdots, MN-1$. This means that one can shift the MSB (most significant bit) of $V(i)$ to $W(i)$ without affecting the encryption result.

\item Differential attack on IEALM

Given two plain-images $\bm{I}_0$ and $\bm{I}_1$ encrypted by
IEALM with the same controlling sequences, we define the bitwise XOR operation between them as $\bm{I}_{0\oplus1}=\bm{I}_0\oplus\bm{I}_1$.
Likewise, $\bm{L}_{0\oplus1}=\bm{L}_0\oplus\bm{L}_1$ and $\bm{H}_{0\oplus1}=\bm{H}_0\oplus\bm{H}_1$, where $(\bm{L}_0, \bm{H}_0)=\Spl(\bm{I}_0)$ and $(\bm{L}_1, \bm{H}_1)=\Spl(\bm{I}_1)$.
As for the corresponding cipher-images $\bm{I}''_0$ and $\bm{I}''_1$, one can get $(\bm{L}_0'', \bm{H}_0'')=\Spl(\bm{I}''_0)$ and $(\bm{L}_1'', \bm{H}_1'')=\Spl(\bm{I}''_1)$.

According to Eq.~\eqref{eq:P*}, one can obtain
\begin{equation}
\label{eq:diffP*}
\left\{
\begin{aligned}
\bm{L}_{0\oplus 1}^* &= (\bm{L}_0\boxplus \bm{V}_L)\oplus(\bm{L}_1\boxplus \bm{V}_L), \\
\bm{H}_{0\oplus 1}^* &= (\bm{H}_0\boxplus \bm{V}_H\boxplus \bm{r}_0)\oplus(\bm{H}_1\boxplus \bm{V}_H\boxplus \bm{r}_1), 
\end{aligned}
\right.
\end{equation}
where $\bm{r}_j=\lfloor(\bm{L}_j'+\bm{V}_L)/16\rfloor$, $j=0, 1$.
Analyzing the above equation individually on each bit-plane, one can get
\begin{equation}
\label{eq:bitleveldiffL*}
L_{0\oplus 1, k}^*(i)=
\begin{cases}
 L_{0\oplus 1, 0}(i)         & \mbox{when } k=0;\\
 L_{0\oplus 1, k}(i)\oplus \theta_{0\oplus 1, k}(i) & \mbox{when }k=1, 2, 3, 
\end{cases}
\end{equation}
and
\begin{equation}
\label{eq:bitleveldiffH*}
H_{0\oplus 1, k}^*(i)=
 \begin{cases}
 H_{0\oplus 1, 0}(i)\oplus r_{0\oplus 1}(i)   & \mbox{when }k=0; \\
 H_{0\oplus 1, k}(i)\oplus \lambda_{0\oplus 1, k}(i) & \mbox{when }k=1, 2, 3, 
 \end{cases}
\end{equation}
where
\begin{equation}
\label{eq:plainpar}
\left\{
\begin{aligned}
 \theta_{j, k}(i)&=\left\lfloor\frac{\sum_{t=0}^{k-1}(V_{L, t}(i)+L_{j, t}(i))\cdot 2^t}{2^k}\right\rfloor, \\
 \lambda_{j, k}(i)&=\left\lfloor\frac{\sum_{t=0}^{k-1}(V_{H, t}(i)+H_{j, t}(i))\cdot 2^t+r_j(i)}{2^k}\right\rfloor, 
\end{aligned}
\right.
\end{equation}
and $j=0, 1$.

According to Eq.~\eqref{eq:newL'}, by differentiating plain-images, one has
\begin{align*}
 L_{0\oplus 1}'(i)&=L_{0\oplus 1}^{\star}(i) \\
 &=\widetilde{L}_{0\oplus 1}^*(i) \oplus \widetilde{H}_{0\oplus 1}^*(i).
\end{align*}
Then in the $k$-bit-plane, one can get 
\begin{align}
\label{eq:bitleveldiffL'}
L_{0\oplus 1, k}'(i) &= L_{0\oplus 1, k}^{\star}(i) \notag \\
      &=L_{0\oplus 1, k}^*(T_{1, k}(i)) \oplus H_{0\oplus 1, k}^*(T_{2, k}(i)).
\end{align}
Similarly, from Eq.~\eqref{eq:newH'} one can get
\begin{align}
\label{eq:bitleveldiffH'}
H_{0\oplus 1, k}'(i) &= H_{0\oplus 1, k}^{\star}(i) \notag \\
&=L_{0\oplus 1, k}^{\star}(T_{3, k}(i))\oplus H_{0\oplus 1, k}^*(T_{4, k}(i))
\end{align}

According to Eq.~\eqref{eq:C}, one has
\begin{equation*}
\left\{
 \begin{aligned}
 \bm{L}_{0\oplus 1}'' &= (\bm{L}_0'\boxplus \bm{V}_L')\oplus(\bm{L}_1'\boxplus \bm{V}_L'), \\
 \bm{H}_{0\oplus 1}'' &= (\bm{H}_0'\boxplus \bm{V}_H'\boxplus \bm{r}_0')\oplus(\bm{H}_1'\boxplus \bm{V}_H'\boxplus \bm{r}_1'), 
 \end{aligned}
\right.
\end{equation*}
where $\bm{r}_j'=\lfloor(\bm{L}_j'+\bm{V}_L')/16\rfloor$ and $j=0, 1$.
Then one can get
\begin{equation}
\label{eq:bitleveldiffL''}
L_{0\oplus 1, k}''(i)=
\begin{cases}
 L_{0\oplus 1, 0}'(i)              & \mbox{when }k=0; \\
 L_{0\oplus 1, k}'(i)\oplus \theta_{0\oplus 1, k}'(i) & \mbox{when }k=1, 2, 3, 
\end{cases}
\end{equation}
and
\begin{equation}
\label{eq:bitleveldiffH''}
H_{0\oplus 1, k}''(i)=
\begin{cases}
 H_{0\oplus 1, 0}'(i)\oplus r_{0\oplus 1}'(i)     &\mbox{when }k=0; \\
 H_{0\oplus 1, k}'(i)\oplus \lambda_{0\oplus 1, k}'(i)  &\mbox{when }k=1, 2, 3, 
\end{cases}
\end{equation}
where
\begin{equation}
\label{eq:chiperpar}
\left\{
\begin{aligned}
 \theta_{j, k}'(i)&=\left\lfloor\frac{\sum_{t=0}^{k-1}(V_{L, t}'(i)+L_{j, t}'(i))\cdot 2^t}{2^k}\right\rfloor, \\
 \lambda_{j, k}'(i)&=\left\lfloor\frac{\sum_{t=0}^{k-1}(V_{H, t}'(i)+H_{j, t}'(i))\cdot 2^t+r'_j(i)}{2^k}\right\rfloor, 
\end{aligned}
\right.
\end{equation}
and $j=0, 1$.

\item Determining $\bm{T}_2$

The bit-level permutation using $\bm{T}_{2, k}$ can be determined with Eq.~\eqref{eq:bitleveldiffH*}, \eqref{eq:bitleveldiffL'}, and \eqref{eq:bitleveldiffL''}.
When $k=0$, one just needs to eliminate $L^*_{0\oplus1}(i)$ and $r_{0\oplus1}(i)$ to attack the permutation.
Through setting $L_{0\oplus1}(i)=0$, one gets $r_{0\oplus1}(i)=0$ and $L^*_{0\oplus1}(i)=0$, and then Eq.~\eqref{eq:bitleveldiffL'} becomes $L'_{0\oplus 1, k}(i)=H^*_{0\oplus 1, k}(T_{2, k}(i))$.

When $k>0$, one needs to additionally eliminate the effect of the carries from the lower bit-planes, namely $\lambda_{0\oplus1, k}(i)$ and $\theta'_{0\oplus1, k}(i)$.
Hence, for $k'=0\sim(k-1)$, set $H_{0\oplus1, k'}(i)=0$, and then $\lambda_{0\oplus1, k}(i)=0$ according to Eq.~\eqref{eq:plainpar}.
Furthermore, one can deduce $L'_{0\oplus1, k'}(i)=0$ and $\theta'_{0\oplus1, k}(i)=0$ from Eq.~\eqref{eq:chiperpar}.
Based on these setting, for $k=0\sim3$, Eq.~\eqref{eq:bitleveldiffL''} becomes
\begin{align*}
L_{0\oplus 1, k}''(i) & = L_{0\oplus 1, k}'(i) \notag\\
       & = H_{0\oplus 1, k}(T_{2, k}(i)).
\end{align*}

In this case, the encryption operation on $H_{0\oplus 1, k}(i)$ is actually permutation-only. As the chosen-plaintext attack on permutation-only encryption algorithm is already mature \cite{Cqli:Scramble:IM17}, we only briefly describe the attack procedure.
The permutation vector $\bm{T}_{2, k}$ can be exactly recovered by the following steps:

\begin{itemize}[leftmargin=*]
\item \textit{Step 1}: Choose a plain-image $\bm{I}_0$ of fixed value zero and get its corresponding cipher-image $\bm{I}_0''$.

\item \textit{Step 2}: Choose $n$ plain-images $\bm{I}_1, \bm{I}_2, \cdots, \bm{I}_n$ that satisfy $\bm{I}_t=\{H_t(i)\cdot 16\}_{i=0}^{MN-1}$, $t=1, 2, \cdots, n$, and the $k'$-th 
significant bit of $H_t(i)$ is
\begin{equation*}
H_{t, k'}(i)=
 \begin{cases}
 \left\lfloor i/2^{t-1}\right\rfloor\bmod2 & \mbox{if } k'=k;\\
 0           & \mbox{otherwise}, 
 \end{cases}
\end{equation*}
where $n=\lceil\log_2({MN})\rceil$ and $k'=0\sim3$.

\item \textit{Step 3}: Encrypt $\bm{I}_1, \bm{I}_2, \cdots, \bm{I}_n$ and get the corresponding cipher-images $\bm{I}_1'', \bm{I}_2'', \cdots, \bm{I}_n''$.
Then, for $\sum_{t=1}^{n}L_{0\oplus t, k}''(i')\cdot2^{t-1}=i$, one can confirm $T_{2, k}(i')=i$, where $L_t''(i')=I_t''(i')\bmod 16$.
\end{itemize}

Perform the above procedures for $k=0\sim3$ to recover $\bm{T}_{2, 0}$, $\bm{T}_{2, 1}$, $\bm{T}_{2, 2}$, and $\bm{T}_{2, 3}$.
In fact, since $\bm{T}_{2, 1}=\bm{T}_{2, 0}$, one can recover $\bm{T}_{2}$ with only $\lceil\log_2({MN})\rceil+1$ plain-images.

\item Determining $\bm{V}_L$

Let $H_{0\oplus 1}(i)=0$, $L_0(i)=0$ and $L_1(i)=c$ for $i=0, 1, \cdots, MN-1$, where $1\leq c\leq15$.
For $k=0$, Eq.~\eqref{eq:bitleveldiffL''} can be expressed as
\begin{align*}
L''_{0\oplus 1, 0}(i)
& = L'_{0\oplus 1, 0}(i) \\
& = L^*_{0\oplus 1, 0}(T_{1, 0}(i)) \oplus H^*_{0\oplus 1, 0}(T_{2, 0}(i)) \\
& = L_{0\oplus 1, 0}(T_{1, 0}(i)) \oplus r_{0\oplus1}(T_{2, 0}(i)) \\
& = (c\bmod 2) \oplus r_1(T_{2, 0}(i)).
\end{align*}
As $r_1(T_{2, 0}(i))=\lfloor(c+V_L(T_{2, 0}(i)))/16\rfloor$, $r_1(T_{2, 0}(i))=1$ if and only if $c+V_L(T_{2, 0}(i))\geq 16$.
Therefore, one can obtain
\begin{equation*}
 L''_{0\oplus1, 0}(i)=
 \begin{cases}
  (c\bmod 2)\oplus 1 & \mbox{if } (c+V_L(T_{2, 0}(i)))\geq 16;\\
  c\bmod 2   & \mbox{otherwise}.
 \end{cases}
\end{equation*}
Setting $c=1$, one can deduce that
\begin{equation*}
V_L(T_{2, 0}(i))\in
\begin{cases}
\{15\} & \mbox{if } L''_{0\oplus1, 0}(i)=0; \\
\{0, 1, \cdots, 14\} & \mbox{otherwise}.
\end{cases}
\end{equation*}
Namely, one can determine $V_L(T_{2, 0}(i))=15$ by finding $L''_{0\oplus1, 0}(i)=0$.
Similarly, one can set $c=2$ and then find $V_L(T_{2, 0}(i))=14$ through
\begin{equation*}
V_L(T_{2, 0}(i))\in
\begin{cases}
\{14, 15\} & \mbox{if } L''_{0\oplus1, 0}(i)=1; \\
\{0, 1, \cdots, 13\} & \mbox{otherwise}.
\end{cases}
\end{equation*}
The remaining unknown elements of $\bm{V}_L$ can be obtained by choosing $c\in \{3, 4, \cdots, 15\}$ in turn.

\item Determining $\bm{T}_1$

The permutation vector $\bm{T}_{1, k}$ can be recovered using Eq.~\eqref{eq:bitleveldiffL*}, \eqref{eq:bitleveldiffL'}, and \eqref{eq:bitleveldiffL''}.
To eliminate $H^*_{0\oplus 1, k}(T_{2, k}(i))$ in Eq.~\eqref{eq:bitleveldiffL'}, one can set $H_0(i)=r_1(i)$ and $H_1(i)=r_0(i)$ to get $H^*_{0\oplus1}(i)=0$.
When $k>0$, choosing $L_{0\oplus1, k'}(i)=0$ for $k'=0\sim(k-1)$, one can deduce $\theta_{0\oplus1, k}(i)=0$ from Eq.~\eqref{eq:plainpar}.
Furthermore, because $L'_{0\oplus1, k'}(i)=L^*_{0\oplus1, k'}(T_{1, k'}(i))=L_{0\oplus1, k'}(i)=0$, the carry $\theta'_{0\oplus1, k}(i)=0$ according to Eq.~\eqref{eq:chiperpar}.
Then, Eq.~\eqref{eq:bitleveldiffL''} becomes
\begin{align*}
L_{0\oplus 1, k}''(i) & = L_{0\oplus 1, k}'(i) \\
            & = L_{0\oplus 1, k}^*(T_{1, k}(i)) \\
            & = L_{0\oplus 1, k}(T_{1, k}(i)).
\end{align*}
Therefore, $\bm{T}_1$ can be exactly determined by utilizing the recovery procedure of $\bm{T}_2$.

\item Determining $\bm{V}_H$

The three LSB of $\bm{V}_H$ can be guessed by observing the corresponding carry 
incurred by the addition in Eq.~\eqref{eq:diffP*}.
Setting $I_0(i)=0$, one has $L_0^*(i)=V_L(i)$ and $r_0(i)=0$.
Since $\bm{V}_L$ is known, one can directly choose $L_1^*(i)=L_0^*(i)\oplus 2^k$ and $H_1(i)=2^k\boxminus r_1(i)$ to determine $\bm{V}_{H, k}$, where $k<3$ and $a\boxminus b=(a-b)\bmod 2^{n0}$.
Note that $r_1(i)$ is also known and $L_1(i)=L_1^*(i)\boxminus V_L(i)$.
Then Eq.~\eqref{eq:diffP*} becomes
\begin{equation*}
H_{0\oplus 1}^*(i)=V_H(i)\oplus(2^k\boxplus V_H(i)).
\end{equation*}
Then one can deduce $L_{0\oplus 1, k}^*(i)=H_{0\oplus 1, k}^*(i)=1$.
When $k>0$, for lower bit-planes, one has $L_{0\oplus 1, k'}^*(i)=H_{0\oplus 1, k'}^*(i)=0$, where $k'=0\sim(k-1)$.
From Eq.~\eqref{eq:bitleveldiffL'}, one can get $L_{0\oplus 1, k'}'(i)=0$ and $L_{0\oplus 1, k}'(i)=0$.
Furthermore, the carry $\theta_{0\oplus 1, k+1}'(i)=0$ according to Eq.~\eqref{eq:chiperpar}.
Hence, Eq.~\eqref{eq:bitleveldiffL''} becomes
\begin{align*}
L_{0 \oplus 1, k+1}''(i)
& = L_{0\oplus 1, k+1}'(i) \\
& = L^*_{0\oplus 1, k+1}(T_{1,k+1}(i))\oplus H^*_{0\oplus 1, k+1}(T_{2,k+1}(i)).
\end{align*}
Considering $L^*_{0\oplus 1, k+1}(T_{1,k+1}(i))=0$, $H_{0\oplus 1, k+1}(T_{2,k+1}(i))=0$, and $\lambda_{0,k+1}(T_{2,k+1}(i))=0$, the above equation can further become
\begin{equation*}
L_{0 \oplus 1, k+1}''(i) = \lambda_{1, k+1}(T_{2, k+1}(i)).
\end{equation*}

Incorporating $\sum_{t=0}^{k}H_{1,t}\cdot2^t+r_1=2^k$ into Eq.~\eqref{eq:plainpar}, one can get
\begin{align*}
\lambda_{1, k+1}(i) & = \left\lfloor\frac{\sum_{t=0}^{k}V_{H, t}(i)\cdot 2^t+ 2^{k}}{2^{k+1}}\right\rfloor \notag \\
         & = \label{eq:prolambda}
        \begin{cases}
          0 & \mbox{when } V_{H, k}(i)=0;\\
          1 & \mbox{when } V_{H, k}(i)=1, 
        \end{cases}
\end{align*}
Namely, $\lambda_{1, k+1}(i)=V_{H, k}(i)$.
Combining the above two equations, one has
\begin{equation*}
V_{H, k}(T_{2, k+1}(i)) = L_{0 \oplus 1, k+1}''(i).
\end{equation*}
Then one can recover $V_{H, 0}(i)$, $V_{H, 1}(i)$, $V_{H, 2}(i)$ by setting $k=0, 1, 2$ in turn.
Shifting $V_{H, 3}(i)$ to $W_{H, 3}(i)$ directly, one can set $V_{H, 3}(i)=0$ according to Fact~\ref{fact:modxorhighbit}.

Since the equivalent version of $\bm{V}$ is known, one can convert any $\bm{I}^*$ to $\bm{I}$ via $\bm{I}=\bm{I}^*\boxminus \bm{V}$.
Therefore, we only consider choosing $\bm{I}^*$ without specifying the plain-image $\bm{I}$ in subsequent discussion.

\item Determining $\bm{T}_4$

The permutation vector $\bm{T}_{4, k}$ can be determined by Eq.~\eqref{eq:bitleveldiffH'} and \eqref{eq:bitleveldiffH''}.
Set $I_0^*(i)=0$ and $\widetilde{L}_1^*(i)=\widetilde{H}_1^*(i)$ to get $L_{0\oplus 1}'(i)=L_{0\oplus 1}^{\star}(i)=0$ and $r_0'(i)=r_1'(i)$.
From Eq.~\eqref{eq:bitleveldiffH'}, one can know that $H_{0\oplus 1, k}'(i)=H^*_{0\oplus 1, k}(T_{4, k}(i))$.
When $k>0$, set $H_{1, k'}^*(i)=0$ for $k'=0\sim(k-1)$ to eliminate the effect of carry $\lambda_{0\oplus 1, k}'(i)$.
Then one can get $H_{0\oplus1, k'}'(i)=0$ and then $\lambda_{0\oplus 1, k}'(i)=0$ from Eq.~\eqref{eq:chiperpar}.
Hence, for $k=0\sim3$, Eq.~\eqref{eq:bitleveldiffH''} becomes
\begin{align*}
H_{0\oplus 1, k}''(i)  & = H'_{0\oplus 1, k}(i) \notag \\
            & = H^*_{0\oplus 1, k}(T_{4, k}(i)).
\end{align*}
Likewise, one can recover $\bm{T}_4$ exactly by utilizing the recovery procedure of $\bm{T}_2$.

\begin{figure*}[!htb]
\centering
\begin{minipage}[t]{\SixImW}
	\centering
	\includegraphics[width=\SixImW]{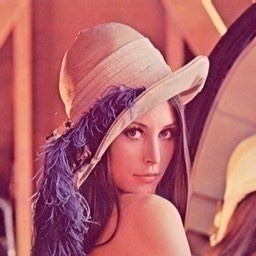}
	a)
\end{minipage}
\begin{minipage}[t]{\SixImW}
	\centering
	\includegraphics[width=\SixImW]{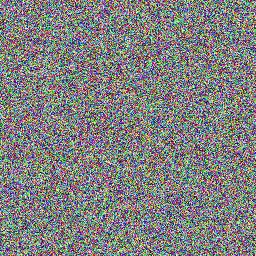}
	b)
\end{minipage}
\begin{minipage}[t]{\SixImW}
	\centering
	\includegraphics[width=\SixImW]{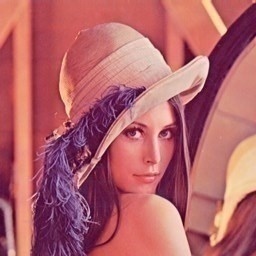}
	c)
\end{minipage}
\begin{minipage}[t]{\SixImW}
	\centering
	\includegraphics[width=\SixImW]{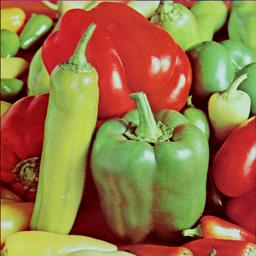}
	d)
\end{minipage}
\begin{minipage}[t]{\SixImW}
	\centering
	\includegraphics[width=\SixImW]{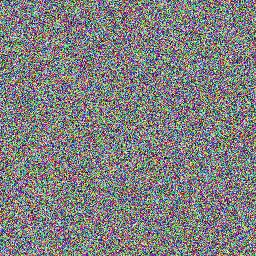}
	e)
\end{minipage}
\begin{minipage}[t]{\SixImW}
	\centering
	\includegraphics[width=\SixImW]{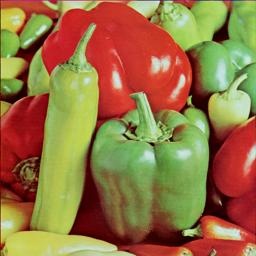}
	f)
\end{minipage}
\caption{Chosen-plaintext attack results on IEALM: a) plain-image ``Lenna"; b) cipher-image of Fig.~a) encryped by $(a, b, x_r, y_g, z_b)=(2, 1.99, 29676, 9202, 62299)$; c) the recovered plain-image with the obtained equivalent secret key; d) plain-image ``Peppers"; e) cipher-image of Fig.~d) encrypted by $(a, b, x_r, y_g, z_b)=(2, 1.99, 29232, 54749, 57603)$; f) the recovered plain-image with the obtained equivalent secret key.}
\label{fig:AttRes}
\end{figure*}

\item Determining $\bm{T}_3$

Let $L_0^*(i)=H_0^*(i)=H_1^*(i)=0$ for $i=0, 1, \cdots, (MN-1)$.
From $L^\star(i)=\widetilde{L}^*(i)\oplus\widetilde{H}^*(i)$ and 
$H^\star(i)=\widehat{L}^\star(i)\oplus\widehat{H}^*(i)$, one can get
$L_0^{\star}(i)=0$, $L_1^{\star}(i)=\widetilde{L}_1^*(i)$, $H_0^{\star}(i)=0$, and $H^\star_{1,k}(i)=L^\star_{1,k}(T_{3,k}(i))$.
Since $L_1^{\star}(i)=\widetilde{L}_1^*(i)$ and $\bm{T}_1$ is known, we directly choose $\bm{L}_1^{\star}$ without specifying the corresponding plain-image in the succeeding discussion.

To determine $\bm{T}_{3,0}$, according to Eq.~\eqref{eq:bitleveldiffH'} and \eqref{eq:bitleveldiffH''}, one has
\begin{align}
\label{eq:atts30}
H_{0\oplus 1, 0}''(i)
& = H_{0\oplus 1, 0}'(i) \oplus r_{0\oplus 1}'(i) \notag \\
& = H_{0\oplus 1, 0}^\star(i) \oplus r_{0\oplus 1}'(i) \notag \\
& = L_{1, 0}^{\star}(T_{3, 0}(i))\oplus r_{0\oplus 1}'(i), 
\end{align}
where $r_{0\oplus 1}'(i)=\lfloor(\beta(i) +V_L'(i))/16\rfloor \oplus \lfloor((\beta(i) \oplus L_1^{\star}(i))+V_L'(i))/16\rfloor$.
Obviously, $r_{0\oplus 1}'(i)$ is only dependent on $L_1^{\star}(i)$, so we use $\Phi(i,L_1^{\star}(i))$ to represent $r_{0\oplus 1}'(i)$ corresponding to different $L_1^{\star}(i)$.
Then, Eq.~\eqref{eq:atts30} can be represented as
\begin{equation*}
H_{0\oplus 1, 0}''(i)\oplus\Phi(i, L_1^{\star}(i))=L_{1, 0}^{\star}(T_{3, 0}(i)).
\end{equation*}
For $L^\star_1(i)=0$, one can see $\Phi(i,0)=0$.
To get $\Phi(i,1)$, one can set $L^\star_1(i)=1$ for $i=0\sim(MN-1)$ and then $\Phi(i,1)=H''_{0\oplus1,0}(i)\oplus1$ from the above equation.
Choosing $L_1^{\star}(i)\in\{0, 1\}$, since $\Phi(i, L_1^{\star}(i))$ is also known, one can recover $\bm{T}_{3, 0}$ with the preceding equation referring to the recovery process of $\bm{T}_{2,k}$.

To determine $T_{3,k}$ for $k\in\{1, 2, 3\}$, likewise, one has
\begin{align}
\label{eq:atts3k}
H_{0\oplus 1, k}''(i) & = H_{0\oplus 1, k}'(i) \oplus \lambda_{0\oplus 1, k}'(i) \notag \\
           & = L_{1, k}^{\star}(T_{3, k}(i))\oplus \lambda_{0\oplus 1, k}'(i).
\end{align}

From Eq.~\eqref{eq:newH'}, one can get $H'_0(i)=\beta'(i)$. Choosing $L^\star_1(i)\in\{0,2^k\}$, for $t=0\sim(k-1)$, $H^\star_{1,t}(i)=L^\star_{1,t}(T_{3,t}(i))=0$ and then $H'_{1,t}(i)=\beta'_{t}(i)$.
According to Eq.~\eqref{eq:chiperpar}, one has
\begin{align*}
\lambda_{0\oplus 1, k}'(i)
&=\left\lfloor\frac{\alpha_k(i)+r_0'(i)}{2^k}\right\rfloor \oplus \left\lfloor\frac{\alpha_k(i)+r_1'(i)}{2^k}\right\rfloor \\
&=\left\lfloor\frac{\alpha_k(i)+\lfloor(\beta(i) +V_L'(i))/16\rfloor}{2^k}\right\rfloor \\
&\oplus \left\lfloor\frac{\alpha_k(i)+\lfloor(\beta(i) \oplus L_1^{\star}(i)+V_L'(i))/16\rfloor}{2^k}\right\rfloor, 
\end{align*}
where $\alpha_k(i)=\sum_{t=0}^{k-1}(V_{H, t}'(i)+\beta_t'(i))\cdot2^t$.
Likewise, we use $\Psi(i, L_1^{\star}(i))$ to represent $\lambda_{0\oplus 1, k}'(i)$ corresponding to different $L^\star_1(i)$.
Then Eq.~\eqref{eq:atts3k} can be expressed as
\begin{equation*}
H_{0\oplus 1, k}''(i)=L_{1, k}^{\star}(T_{3, k}(i))\oplus \Psi(i, L_1^{\star}(i)).
\end{equation*}
For $L^\star_1(i)=0$, one can know $\Psi(i, 0)=0$.
To obtain $\Psi(i, 2^k)$, set $L^\star_1(i)=2^k$ for $i=0\sim(MN-1)$ and then get $\Psi(i, L_1^{\star}(i))=H_{0\oplus 1, k}''(i)\oplus1$ from the preceding equation.
Choosing $L^\star_1(i)\in\{0,2^k\}$, since $\Psi(i, L_1^{\star}(i))$ is known, one can recover $\bm{T}_{3,k}$ using the above equation and the recovery process of $\bm{T}_{2,k}$.
Therefore, $\bm{T}_3=\left\{\bm{T}_{3, 0}, \bm{T}_{3, 1}, \bm{T}_{3, 2}, \bm{T}_{3, 3}\right\}$ is determined.

\begin{table*}[!htb]
\centering
\caption{Some data in the encryption process of the blue channel of Fig.~\ref{fig:AttRes}b).}
\begin{tabular}{c|*{11}{c}}
\hline
\diagbox{Item}{Index} & 1 & 2 & 4 & 8 & 16 & 32 & 64 & 128 & 256 & 512 & 1024\\
\hline
$I(i)$    & 125  & 123  & 122  & 114  & 110  & 83  & 100  & 106  & 125  & 122  & 114\\
$T^{2, 0}(i)$ & 63654 & 41166 & 44389 & 5418 & 60541 & 8324 & 8394 & 52758 & 10693 & 18236 & 12940\\
$T^{1, 1}(i)$ & 62246 & 12618 & 22576 & 424  & 5892 & 47186 & 18568 & 14185 & 4948 & 47571 & 6740 \\
$T^{4, 2}(i)$ & 8436 & 37177 & 13122 & 24285 & 25840 & 24911 & 350  & 52730 & 12436 & 30075 & 132\\
$T^{3, 3}(i)$ & 1357 & 27981 & 60186 & 16982 & 691  & 9877 & 32352 & 30284 & 62723 & 61986 & 27694\\
$V(i)$    & 72  & 61  & 201  & 128  & 210  & 239  & 54  & 92  & 42  & 22  & 199\\
$I^*(i)$   & 197  & 184  & 67  & 242  & 64  & {66}  & 154  & 198  & 167  & 144  & {57}\\
$I'(i)$   & 241  & 191  & 78  & 14  & 227  & 4   & 172  & 169  & 53  & 31  & 252\\
$I''(i)$    & 207  & 70  & 75  & 7  & 173  & 226  & 225  & 87  & 5   & 190  & 196\\
\hline
\end{tabular}
\label{tab:encdata}
\end{table*}

\begin{table*}[!htb]
\centering
\caption{The items obtain in an attack corresponding to the data shown in Table~\ref{tab:encdata}.}
\begin{tabular}{c|*{11}{c}}
\hline
\diagbox{Item}{Index} & 1 & 2 & 4 & 8 & 16 & 32 & 64 & 128 & 256 & 512 & 1024\\
\hline
$T^{2, 0}(i)$ & 63654 & 41166 & 44389 & 5418 & 60541 & 8324 & 8394 & 52758 & 10693 & 18236 & 12940\\
$T^{1, 1}(i)$ & 62246 & 12618 & 22576 & 424  & 5892 & 47186 & 18568 & 14185 & 4948 & 47571 & 6740 \\
$T^{4, 2}(i)$ & 8436 & 37177 & 13122 & 24285 & 25840 & 24911 & 350  & 52730 & 12436 & 30075 & 132\\
$T^{3, 3}(i)$ & 1357 & 27981 & 60186 & 16982 & 691  & 9877 & 32352 & 30284 & 62723 & 61986 & 27694\\
$V(i)$    & 72  & 61  & \underline{73}  & \underline{0}   & \underline{82}  & \underline{111}  & 54  & 92  & 42  & 22  & \underline{71}\\
$I^*(i)$   & 197  & 184  & \underline{195}  & \underline{114}  & \underline{192}  & \underline{194}  & 154  & 198  & 167  & 144  & \underline{185}\\
$I(i)$    & 125  & 123  & 122  & 114  & 110  & 83  & 100  & 106  & 125  & 122  & 114\\
\hline
\end{tabular}
\label{tab:attdata}
\end{table*}

\item Attacking the other encryption operations

Once all permutation vectors were successfully recovered, one can get
\begin{align*}
I''(i) & = W'(i)\oplus (I'(i)\boxplus V'(i)) \\
    & = W'(i)\oplus (((\beta(i)+16\cdot \beta'(i))\oplus I^{\star}(i))\boxplus V'(i))
\end{align*}
from Eq.~\eqref{eq:C}, \eqref{eq:newL'} and \eqref{eq:newH'}, where $I^{\star}(i)=L^{\star}(i)+ H^{\star}(i)\cdot 16$.
In the above equation, $I''(i)$ can be determined by $I^{\star}(i)$ and vice versa.
In other words, $I''(i)$ and $I^{\star}(i)$ are one-to-one correspondence.
Note that one can directly choose any $\bm{I}^{\star}$ and then obtain its plain-image $\bm{I}$.
Accordingly, choose $I^{\star}(i)=c$ for $c=0\sim255$, and then get $I''(i)$ to obtain a map $F(i,I''(i))=c$, which can be used to recover $I^{\star}(i)$ from $I''(i)$.

\item Recovering the plain-image

As for a cipher-image $\bm{I}''$, first recover $\bm{I}^{\star}$ by $I^{\star}(i)=F(i,I''(i))$ and calculate $(\bm{L}^{\star}, \bm{H}^{\star})=\Spl(\bm{I}^{\star})$.
Perform the permutation with $\bm{T}_3$ on $\bm{L}^{\star}$ to get $\widehat{\bm{L}}^{\star}$, and derive $\widehat{\bm{H}}^*$ by $\widehat{H}^*(i)=H^{\star}(i)\oplus \widehat{L}^{\star}(i)$.
Apply the inverse version of permutation with $\bm{T}_4$ on $\widehat{\bm{H}}^*$ to recover $\bm{H}^*$.
Permute $\bm{H}^*$ with $\bm{T}_2$ to get $\widetilde{\bm{H}}^*$, and derive $\widetilde{\bm{L}}^*$ by $\widetilde{L}^*(i)=L^{\star}(i)\oplus \widetilde{H}^*(i)$.
Then employ the inverse version of permutation with $\bm{T}_1$ on $\widetilde{\bm{L}}^*$ to recover $\bm{L}^*$.
Finally, one can get plain-image $\bm{I}$ by $I(i)=(L^*(i)+16\cdot H^*(i))\boxminus V(i)$.
\end{itemize}

To verify the effectiveness of the above attacking process, we performed a number of experiments with some random secret keys
\footnote{The source codes of this paper are available at \url{https://github.com/ChengqingLi/MM-IEALM}}. The attack results on a typical image is shown in Fig.~\ref{fig:AttRes}. The success rate of the attack is 100\%. And some
recovered data during the attack are listed in Tables~\ref{tab:encdata} and \ref{tab:attdata} as a reference. Note that the recovered MSB of $V(i)$ and $I^*(i)$
may be different from that in Table~\ref{tab:encdata} (the underlined elements) as it has no influence on the decryption result.

\subsection{Complexity of the attack}

Recalling the attack process, consider $\bm{T}_{1, 0}=\bm{T}_{1, 1}$, $\bm{T}_{2, 0}=\bm{T}_{2, 1}$, $\bm{T}_{3, 0}=\bm{T}_{3, 1}$ and $\bm{T}_{4, 0}=\bm{T}_{4, 1}$, one can see that the images required in procedure determining $\bm{T}_2$ and $\bm{T}_4$ are $P_1=P_2=\lceil\log_2({MN})\rceil+1$. And the number of images needed to recover $\bm{V}$ is $P_3=6$. Likewise, one can get determining $\bm{T}_1$ and $\bm{T}_3$ need $P_4=2\cdot\lceil\log_2({MN})\rceil$ and $P_5=\lceil\log_2({MN})\rceil+1$ images, respectively. The final procedure of attack requires $P_6=86$ images. Since the complexity of each procedure is $O(P_{i}MN)$, the complexity of the attack is $O(PMN)\approx O(MN\log_2(MN))$, where $P=P_1+P_2+\cdots+P_6$. 

As for the image shown in Fig.~\ref{fig:AttRes} c), the numbers of chosen plain-images required for the above attacking procedures are 17, 17, 6, 32, 17, and 86, respectively. The complexity of the attack is $O(175MN)$, this attack takes only a few seconds on a personal computer. 

\subsection{Analysis of the key space}

The designers of IEALM claimed that its key space is greater than $10^{120}$ \cite{zhangf:2Dlag:IM21}.
In fact, the real key space of IEALM depends on arithmetic precision $L$ and the pixel values of the plain-image. Although the pseudorandom number generator has six initial values, they are only determined by the pixel values of the plain-image. This makes the actual key space of IEALM can not reach that estimated by the designers.
Strictly speaking, the key space of IEALM is only $2^{2L}\cdot (256MN)^3$. As for a plain-image of size $2048\times 2048$, the sizes of key space of IEALM implemented with precision 32-bit and 64-bit are $10^{46}$ and $10^{65}$, respectively.
In contrast, for a plain-image of size $256\times 256$, the sizes are reduced to $10^{38}$ and $10^{57}$, respectively.
In any case, the size of key space is far smaller than $10^{120}$ claimed by the designers.
Note that the above estimation is only the maximum theoretical value of the key space.
As the keys are not uniformly distributed and not all control parameters of a chaotic map can generate sequences with enough randomness, the real effective key space is even much smaller.

\section{Conclusion}

This paper analyzed the security performance of an image encryption algorithm based on 2D lag-complex logistic map from the viewpoint of modern cryptanalysis.
The sensitivity mechanism between the controlling pseudo-random sequences and the plain-image can be cancelled for some special plain-images. Due to the algorithm is composed by cascading some independent basic encryption parts, their equivalents were recovered with some chosen plain-images one by one. The properties of the underlying chaotic map were explored to enhance the attacking efficiency. Both theoretical analysis and experimental results were provided to show the effectiveness of the attack. The presented results show that designing a secure and efficient image encryption algorithm should consider many factors, such as special storage format of image data, randomness of the adopted pseudo-random number generator, application scenarios and real structure of the encryption algorithm.

\section*{Acknowledgements}

This work was supported in part by the National
Natural Science Foundation of China under Grant
61772447 and Scientific Research Fund of Hunan Provincial Education
Department under Grant 20C1759.

\bibliographystyle{IEEEtran_doi}
\bibliography{MM-IELCS}

\end{document}